\documentclass[11pt,a4paper]{article}
\usepackage[]{geometry}                
\usepackage{graphicx}
\usepackage{amssymb,amsmath,amsfonts}
\usepackage{amsthm}
\usepackage{color}
\usepackage{bm}
\usepackage{hyperref}
\usepackage{authblk}
\usepackage{todonotes}
\usepackage{changes}

\newcommand{\yp}[1]{{\color{black}{#1}}}
\newcommand{\ypb}[1]{{\color{black}{#1}}}

\newcommand{\ve}{\varepsilon}
\newcommand{\pa}{\partial}
\newcommand{\vc}{\text{vec}}
\newcommand{\h}{\mathcal{H}}

\title{High-Order Accuracy Computation of Coupling Functions for Strongly Coupled Oscillators}
\author[1]{Youngmin Park\footnote{Corresponding author ypark@brandeis.edu}}
\author[2]{Dan Wilson}

\affil[1]{Department of Mathematics, Brandeis University, Waltham, MA 02453}
\affil[2]{Department of Electrical Engineering and Computer Science, University of Tennessee, Knoxville, TN 37996}

\date{}

\begin{document}

	\maketitle
	\begin{abstract}
		We develop a general framework for identifying phase reduced equations for finite populations of coupled oscillators that is valid far beyond the weak coupling approximation. This strategy represents a general extension of the theory from [Wilson and Ermentrout, Phys.~Rev.~Lett 123, 164101 (2019)] and yields coupling functions that are valid to \yp{higher-order} accuracy in the coupling strength \yp{for arbitrary types of coupling (e.g., diffusive, gap-junction, chemical synaptic)}. These coupling functions can be used to understand the behavior of potentially high-dimensional, nonlinear oscillators in terms of their phase differences. The proposed formulation accurately replicates nonlinear bifurcations that emerge as the coupling strength increases and is valid in regimes well beyond those that can be considered using classic weak coupling assumptions. We demonstrate the performance of our approach through two examples. First, we use \yp{diffusively coupled} complex Ginzburg-Landau (CGL) model and demonstrate that our theory accurately predicts bifurcations far beyond the range of existing coupling theory. Second, we use a realistic conductance-based model of a thalamic neuron and show that our theory correctly predicts asymptotic phase differences for non-weak \yp{synaptic coupling}. In both examples, our theory accurately captures model behaviors that \yp{weak coupling theories} can not.
	\end{abstract}
	
	\normalsize
	
	\section{Introduction}
	Self-sustained oscillations are observed in a wide array of biological \cite{winfree2001geometry,izhikevich2007}, physical \cite{strogatz2005crowd,ott2008low}, and chemical \cite{kuramoto84,epstein1998introduction} systems. A common and powerful approach to understanding how network oscillators interact is the phase reduction method \cite{kuramoto84,izhikevich2007,ermentrout2010,park2017utility}. Its utility comes from reducing a network of $N$ general $n$-dimensional oscillators into $N-1$ equations that characterize the temporal evolution of phase differences. Indeed, the weak coupling paradigm has driven much work on coupled oscillators in recent decades \cite{ermentrout1981,ek84,schwemmer,dorfler2013,park2016weakly,pimenova2016interplay}.
	
	Unfortunately, the weak coupling assumption cannot accurately capture the dynamical behavior of coupled oscillator networks in many practical applications. This limitation is especially true in many biological systems.  For instance, while individual cortical neurons elicit small-magnitude postsynaptic responses \cite{hoppensteadt1997}, the postsynaptic neuron receives tens of thousands of such responses, resulting in effectively strong coupling \cite{pfeffer2013inhibition}. Subcortical networks such as the basal ganglia include strong synaptic conductances \cite{thibeault2013using}. Pacemaker neurons such as those in the pre-Boetzinger complex and crab stomatogastric ganglion have coupling strengths several orders of magnitude beyond the regime for which the weak coupling approximation is valid \cite{butera1999models,golowasch1992contribution}. For weak coupling to serve as a good approximation in these cases, perturbed trajectories must remain within a small neighborhood of the underlying limit cycle -- a particularly restrictive requirement for limit cycles that have slowly decaying transients \cite{wilson2016isostable,ermentrout2019recent}. 
	
	\yp{To overcome the weak coupling assumption, researchers have used particular tractable models such as integrate-and-fire models \cite{van1994inhibition,ermentrout2002modeling}, or used common features in coupled biological oscillators such as pulse-like coupling \cite{cui2009functional, canavier2009phase,canavier2010pulse,peskin1975mathematical,mirollo1990synchronization} to make problems tractable. We wish to establish a general extension of weak coupling theory for potentially high-dimensional oscillators that includes non-pulsatile coupling.
		
		Recent work in this direction includes} \cite{wilson2019phase}, the authors derive a general second-order correction to the classic first-order theory of weakly coupled oscillators. The method exploits the theory of isostable coordinates \cite{maur13, wilson2018greater}, which represent level sets of the slowest decaying modes of the Koopman operator \cite{mezi13,budi12} to derive the higher-order accuracy corrections for the phase dynamics. The authors in \cite{rosenblum2019numerical,genge2020high} introduce a general numerical method to numerically estimate higher-order phase equations. Finally, although not directly a coupling result, the results of \cite{wilson2020phase} are highly relevant, where Wilson introduced a phase reduction method for strong perturbations using isostable coordinates.
	
	In this paper, we develop a general framework that can be used to identify coupling functions that are valid to \yp{arbitrary accuracy using an asymptotic expansion in the reduced order coordinates}. Related work by \cite{rosenblum2019numerical,genge2020high} requires estimations obtained by the phase dynamics over time, i.e., the full model must be computed for potentially long times and become difficult to implement in high dimensions. In contrast, we exploit the higher-order isostable reduction from \cite{wilson2020phase} and derive high-accuracy phase-interaction functions to \yp{higher-order} accuracy in the coupling strength. This work extends upon previous results in \cite{wilson2019phase, wilson2019augmented} that computed second-order accurate coupling functions using the isostable coordinate framework.
	
	By restricting our attention to a hypersurface defined by the slowest decaying isostable coordinates, the resulting framework can be readily implemented even if the underlying models are high-dimensional.  Furthermore, the numerical implementation only involves computing a hierarchy of scalar ODEs and scalar integrals and does not require {\it a priori} knowledge of the phase trajectories. \yp{One caveat is that we use first-order averaging theory. However, we find that first-order averaging is sufficient to capture non-weak coupling dynamics in our examples.}
	
	We organize the paper as follows. In Section \ref{sec:nonweak}, we introduce our general phase reduction method for $N$ coupled oscillators. In Section \ref{sec:nonweak_example}, we demonstrate up to order $\ve^3$ how our symbolic solver derives the reduced equations using $N=2$ oscillators. We apply our results to the complex Ginzburg-Landau (CGL) ODE model in Section \ref{sec:cgl} and a realistic conductance-based neural model of a thalamic neuron in Section \ref{sec:thal}. We conclude with a discussion in Section \ref{sec:discussion}.
	
	All code used to generate the phase equations are publicly available on GitHub at \url{https://github.com/youngmp/strongcoupling}. Our open-source implementation is written in Python \cite{van1995python}. The repository includes documentation on how to use our software for general systems and includes additional examples.
	
	\section{Derivation}\label{sec:nonweak}
	
	In this section, we reduce the dynamics of $N$ strongly coupled oscillators to a system of $N-1$ equations representing the phase differences. We begin with the autonomous ODEs
	\begin{equation}\label{eq:odes}
		\dot X_i = F(X_i) +\ve \sum_{j=1}^N a_{ij} G(X_i,X_j), \quad i=1,\ldots,N,\\
	\end{equation}
	where each system admits a $T$-periodic limit cycle $Y(t)$ when $\ve=0$. We allow $\ve>0$ not necessarily small and assume general smooth vector fields $F:\mathbb{R}^n \rightarrow \mathbb{R}^n$ and a smooth coupling function $G:\mathbb{R}^n\times\mathbb{R}^n\rightarrow \mathbb{R}^n$. The scalars $a_{ij}$ modulate the strength of coupling between pairs of oscillators, whereas $\ve$ modulates the overall coupling strength of the network. Throughout the text, we will use subscripts $i$ and $j$ to denote oscillator indices and superscripts $k$ and $\ell$ to denote exponents and expansions.
	
	Similar to prior studies \cite{wilson2018greater,wilson2019phase}, we make the explicit assumption that all but one of the $n-1$ non-unity Floquet multipliers is sufficiently close to $0$ so that only a single isostable coordinate is required per oscillator. Additional isostable coordinates could be considered with appropriate modifications to the derivation to follow.  Let $\kappa<0$ be the corresponding Floquet exponent. Using the theory of isostable reduction \cite{wilson2019phase,wilson2020phase}, Equation \eqref{eq:odes} reduces to the phase-amplitude coordinates,
	\begin{equation}\label{eq:reduced0}
		\begin{split}
			\dot \theta_i &= 1+\ve\mathcal{Z}(\theta_i,\psi_i) \cdot \sum_{j=1}^N a_{ij} G(\theta_i,\yp{\psi_i},\theta_j,\yp{\psi_j}),\\
			\dot \psi_i &= \kappa \psi_i + \ve\mathcal{I}(\theta_i,\psi_i)\cdot \sum_{j=1}^N a_{ij} G(\theta_i,\yp{\psi_i},\theta_j,\yp{\psi_j}).
		\end{split}
	\end{equation}
	where $\theta_i$ represents the phase of oscillator $i$ and $\psi_i$ represents the amplitude of a trajectory perturbed away from the underlying limit cycle. \yp{Note that $\theta_i$ is a function of time, but we will generally suppress this dependence for notational convenience in the derivation to follow. We will later show that the variable $\psi_i$, once expanded in $\ve$, can be expressed in terms of $\theta_i,\theta_j$, thus reducing the dimension of the system to one.}
	
	The functions $\mathcal{Z}$ and $\mathcal{I}$ can be computed to arbitrarily high accuracy  by computing coefficients of the expansions:
	\begin{align}
		\mathcal{Z}(\theta,\psi) &\approx Z^{(0)}(\theta) + \psi Z^{(1)}(\theta) + \psi^2 Z^{(2)}(\theta) +\ldots,\label{eq:z_exp}\\
		\mathcal{I}(\theta,\psi) &\approx I^{(0)}(\theta) + \psi I^{(1)}(\theta) + \psi^2 I^{(2)}(\theta) +\ldots,\label{eq:i_exp}\\
		X_i(t) &\approx Y(\theta_i) + \psi_i g^{(1)}(\theta_i)+ \psi_i^2g^{(2)}(\theta_i)+\ldots,\label{eq:x_exp}\\
		\psi_i(t) &\approx \ve p_i^{(1)}(t) + \ve^2 p_i^{(2)}(t) + \ve^3 p_i^{(3)}(t) + \ldots\label{eq:psi_exp},
	\end{align}
	where $Z^{(k)}$, $I^{(k)}$, and $g^{(k)}$ are the \yp{phase response curve (PRC), isostable response curve (IRC)}, and Floquet eigenfunction expansions respectively, $\theta_i$ are the phase variables of each oscillator and $\psi_i$ are the amplitude coordinates. Using the method in \cite{wilson2020phase}, these functions can be computed numerically provided the underlying equations are known.  We will assume that we have performed such computations for a given system and have solutions $Z^{(k)}$, $I^{(k)}$, and $g^{(k)}$ for each $k$ (our Python implementation includes methods that automate the computation of these functions).
	
	Next, we expand the coupling function $G$ in powers of $\ve$. Let us fix a particular pair of oscillators $i$ and $j$. To expand $G$ in powers of $\ve$,  we use the Floquet eigenfunction approximation
	\begin{equation}\label{eq:efuns}
		\Delta x_i \approx \psi_i g^{(1)}(\theta_i) + \psi_i^2 g^{(2)}(\theta_i) + \ldots, \end{equation}
	where $\Delta x_i \equiv X_i(t)-Y(\theta_i\yp{(t)})$. We view the coupling function as the map $G:\mathbb{R}^{2n}\rightarrow \mathbb{R}^n$ where $G(X)=\left[G_{1},\ldots G_{n}\right]^T \in\mathbb{R}^n$, $G_{m}:\mathbb{R}^{2n}\rightarrow \mathbb{R}$, and $X = \left[X_i^T,X_j^T\right]^T \in \mathbb{R}^{2n}$. We then apply the standard definition of higher-order derivatives from \cite{magnus2019matrix,wilson2020phase} to obtain the multivariate Taylor expansion in $\Delta x_i$.
	
	Starting with \yp{an arbitrary $m$th coordinate of the vector-valued function} $G(Y+\Delta X)$, where $Y=[Y(\theta_i)^T,Y(\theta_j)^T]^T$, and $\Delta X = [\Delta x_i^T, \Delta x_j^T]^T$ \yp{(both $2n \times 1$ column vectors)}, the Taylor expansion yields
	\begin{align}\label{eq:g_raw}
		G_m(Y + \Delta X)&= G_m(Y)+ G_m^{(1)}(Y)\Delta X+ \sum_{k=2}^\infty \frac{1}{k!}\left[ \stackrel{k}{\otimes} \Delta X^T\right] \vc\left(G_m^{(k)}(Y)\right),
	\end{align}
	where 
	\begin{equation}\label{eq:G0}
		G_m^{(k)} = \frac{\pa \vc\left(G_m^{(k-1)}\right)}{\pa X^T} \in \mathbb{R}^{\yp{(2n)}^{(k-1)}\times \yp{2n}}.
	\end{equation}
	That is, the partial of $G$ is taken with respect to all coordinates of oscillators $i$ and $j$. We replace $\Delta X$ in Equation \eqref{eq:G0} with the Floquet eigenfunction expansions (Equation \eqref{eq:efuns}) and replace each $\psi_i^k$ with the expansion for $\psi_i$ (Equation \eqref{eq:psi_exp}). With these substitutions in place, we collect the expansion of $G$ in powers of $\ve$. In general, the notation becomes cumbersome, so we summarize this step by writing
	\begin{equation}\label{eq:G_eps}
		\begin{split}
			G(\theta_i,\yp{\psi_i},\theta_j,\yp{\psi_j})= & K^{(0)}(\theta_i,\theta_j)\\
			&+ \ve K^{(1)}\left(\theta_i,\theta_j\yp{,p_i^{(1)},p_j^{(1)}}\right)\\
			&+ \ve^2 K^{(2)}\left(\theta_i,\theta_j,\yp{p_i^{(1)},p_i^{(2)},p_j^{(1)},p_j^{(2)}}\right)\\
			&+ \ldots.
		\end{split}
	\end{equation}
	The $O(1)$ $K^{(k)}$ functions are the appropriately-collected terms including partials of $G$ and the Floquet eigenfunctions. In the calculations to follow, we often suppress the dependence on the $O(1)$ functions $p_i^{(k)},p_j^{(k)}$. \yp{We refer the reader to Appendix \ref{a:K} for the details of Equation \eqref{eq:G_eps}}. \yp{It is straightforward to verify (using a symbolic package) that for a given $k$, each $K^{(\ell)}$ term only depends on terms $p_i^{(\ell)}$, $p_j^{(\ell)}$ for $\ell \leq k$.}
	
	At this step, we have all the necessary expansions in $\ve$ to rewrite the phase-amplitude equations in Equation \eqref{eq:reduced0} in powers of $\ve$. However, this system is still in two dimensions per oscillator. In order to reduce the equations to one per oscillator, we solve for $\psi_i$ in terms of $\theta_i,\theta_j$. To this end, we proceed with the method suggested by \cite{wilson2019phase}.
	
	Making the substitution $\hat\theta_i = \theta_i-t$ in Equation \eqref{eq:reduced0} yields,
	\begin{align}
		\dot{\hat\theta}_i &= \ve \sum_{j=1}^N a_{ij} \mathcal{Z}(\hat\theta_i+t,\psi_i) \cdot G(\hat\theta_i+t,\hat\theta_j+t),\label{eq:th_moving}\\
		\dot \psi_i &=\kappa \psi_i + \ve \sum_{j=1}^N a_{ij} \mathcal{I}(\hat\theta_i+t,\psi_i)\cdot G(\hat\theta_i+t,\hat\theta_j+t).\label{eq:psi_moving}
	\end{align}
	Now substituting the expansion $\psi_i(t) = \ve p_i^{(1)}(t) + \ve^2 p_i^{(2)}(t) + \ve^3 p_i^{(3)}(t) + \ldots,$ into Equation \eqref{eq:psi_moving}, yielding a hierarchy of ODEs in powers of $\ve$ of $\psi_i$ in terms of $\hat \theta_i,\hat \theta_j$. The left-hand consists of straightforward time-derivatives:
	\begin{equation*}
		\psi_i' = \ve \frac{d}{dt}p_i^{(1)} + \ve^2 \frac{d}{dt}p_i^{(2)}+ \ve^3 \frac{d}{dt}p_i^{(3)} +\ldots.
	\end{equation*}
	The right-side, after plugging in the function expansions, reads
	\begin{align*}
		\kappa \psi_i + &\ve\sum_{j=1}^N a_{ij}\mathcal{I}(\hat \theta_i+t,\psi_i)\cdot G(\hat\theta_i+t,\hat\theta_j+t) \\
		&= \kappa\left[\ve p_i^{(1)}(t) + \ve^2 p_i^{(2)}(t)+\ldots\right]\\
		&\quad+ \ve\sum_{j=1}^N  a_{ij}\left(\left[ I_i^{(0)}(\hat\theta_i+t) + \psi I_i^{(1)}(\hat\theta_i+t) + \psi^2 I_i^{(2)}(\hat\theta_i+t) +\ldots\right]\right.\\
		&\quad\quad\quad\quad \quad  \left.\cdot\left[ K_i^{(0)}(\hat\theta_i+t,\hat\theta_j+t) + \ve K_i^{(1)}(\hat\theta_i+t,\hat\theta_j+t) + \ve^2 K_i^{(2)}(\hat\theta_i+t,\hat\theta_j+t) + \ldots\right]\right).
	\end{align*}
	These expansions yield the hierarchy of scalar ODEs in $\ve$,
	\begin{align*}
		O(\ve):\quad \frac{dp_i^{(1)}}{dt} &= \kappa p_i^{(1)}(t) + \sum_{j=1}^N  a_{ij} I^{(0)} \cdot K^{(0)},\\
		O(\ve^2):\quad\frac{dp_i^{(2)}}{dt} &= \kappa p_i^{(2)} + \sum_{j=1}^N  a_{ij} \left(I^{(0)} \cdot K^{(1)} + p_i^{(1)} I^{(1)}\cdot K^{(0)}\right),\\
		O(\ve^3):\quad\frac{dp_i^{(3)}}{dt} &= \kappa p_i^{(3)} + \sum_{j=1}^N  a_{ij} \left( I^{(0)} \cdot K^{(2)} + p_i^{(1)}I^{(1)}\cdot K^{(1)}\right.\\
		&\quad\quad\quad\quad\quad\quad\quad+\left. p_i^{(2)}I^{(1)}\cdot K^{(0)}   + \left(p_i^{(1)}\right)^2 I^{(2)}\cdot K^{(0)}\right),\\
		&\vdots
	\end{align*}
	where $p_i^{(k)}$ are functions of time $t$ (with phase shifts in $\hat \theta_i$ and $\hat \theta_j$ as we will show below), $I^{(k)}$ are functions of $\hat\theta_i+t$, and $K^{(k)}$ are functions of $\hat\theta_i+t,\hat\theta_j+t$. Note that all ODEs are first-order inhomogeneous differential equations with forcing terms that depend on lower-order solutions, so we can solve each ODE explicitly. \yp{In particular, the forcing functions $f^{(k)}(\hat\theta_i+t,\hat\theta_j+t)$ are the summed terms above:
		\begin{align*}
			f^{(1)}(\hat\theta_i+t,\hat\theta_j+t) &= \sum_{j=1}^N a_{ij} I^{(0)}\cdot K^{(0)}\\
			f^{(2)}(\hat\theta_i+t,\hat\theta_j+t) &= \sum_{j=1}^N  a_{ij} \left(I^{(0)} \cdot K^{(1)} + p_i^{(1)} I^{(1)}\cdot K^{(0)}\right),\\
			f^{(3)}(\hat\theta_i+t,\hat\theta_j+t) &=\sum_{j=1}^N  a_{ij} \left( I^{(0)} \cdot K^{(2)} + p_i^{(1)}I^{(1)}\cdot K^{(1)}+ p_i^{(2)}I^{(1)}\cdot K^{(0)}   + \left(p_i^{(1)}\right)^2 I^{(2)}\cdot K^{(0)}\right),\\
			&\vdots
		\end{align*}
		The integrating factor method yield a solution for $p_i^{(k)}$ in terms of the forcing function $f^{(k)}$,}
	\begin{equation*}
		p_i^{(k)}(t) = \sum_{j=1}^N  a_{ij}\yp{\int_{t_0}^t} e^{\kappa(t-s)} f^{(k)}(\hat\theta_i+s,\hat\theta_j+s) ds + e^{\kappa t}C, \quad \yp{k}=1,2,\ldots
	\end{equation*}
	where $C$ is a constant of integration. To discard transients, we ignore the constant of integration and \yp{take $t_0 \rightarrow -\infty$}. For convenience, we also make the change of variables $s\rightarrow t-s$. Then the solutions become,
	\begin{align}\label{eq:p_i}
		p_i^{(k)}(t) &=\sum_{j=1}^N  a_{ij} \int_0^{\infty} e^{\kappa s}f^{(k)}(\hat\theta_i+t-s,\hat\theta_j+t-s)ds\\
		&= \tilde p_i^{(k)}(\hat\theta_1+t,\ldots,\hat\theta_N+t).
	\end{align}
	\yp{Note that $p_i^{(k)}(t)$ is a function of time, whereas $\tilde p_i^{(k)}$ is a function of space. In particular, $\tilde p_i^{(k)}$ acts on the $N$-torus}. Recalling that $\yp{\tilde p_i^{(k)}}$ are coefficients of the $\ve$-expansion of $\psi_i$, it follows that each $\psi_i$ \yp{can be written directly in terms of $\hat\theta_1,\ldots,\hat\theta_N$ and we have therefore eliminated the equation for $\psi_i$} (note the lowest order $\yp{\tilde p_i^{(k)}}$ is the same function as the function $r_j$ as \cite{wilson2019phase}).
	
	\yp{We now simplify Equation \eqref{eq:th_moving} by introducing the expansions derived above:
		\begin{align*}
			\dot{\hat\theta}_i &= \ve \sum_{j=1}^N a_{ij} \mathcal{Z} (\hat\theta_i + t,\psi_i) \cdot G(\hat\theta_i+t,\hat\theta_j+t) \\
			&= \ve \sum_{j=1}^N a_{ij} \left[Z^{(0)}+\psi_i Z^{(1)} + \psi_i^2Z^{(2)}+\psi_i^3 Z^{(3)}+\ldots\right]\cdot \left[K^{(0)} + \ve K^{(1)} + \ve^2 K^{(2)} + \ve^3 K^{(3)}+\ldots\right].
		\end{align*}
		Substituting the expansion for $\psi_i$ and collecting in powers of $\ve$ yields,
		\begin{align*}
			\dot{\hat\theta}_i &= \ve \sum_{j=1}^N a_{ij} K^{(0)}\cdot Z^{(0)}\\
			&\quad +\ve^2 \sum_{j=1}^N a_{ij} K^{(1)}\cdot Z^{(0)} + \tilde p_i^{(1)} K^{(0)}\cdot Z^{(1)}\\
			&\quad +\ve^3 \sum_{j=1}^N a_{ij} K^{(2)}\cdot Z^{(0)} + \tilde p_i^{(1)}K^{(1)} \cdot Z^{(1)} + \tilde p_i^{(2)}K^{(0)} \cdot Z^{(1)} + \left(\tilde p_i^{(1)}\right)^2 K^{(0)}\cdot Z^{(2)}\\
			&\vdots
		\end{align*}
		Note that the suppressed dependencies are as follows: $K^{(0)} = K^{(0)}(\hat\theta_i+t,\hat\theta_j+t)$, $K^{(1)} = K^{(1)}(\hat\theta_i+t,\hat\theta_j+t,\tilde p_i^{(1)},\tilde p_j^{(1)})$, $K^{(2)} = K^{(2)}(\hat\theta_i+t,\hat\theta_j+t,\tilde p_i^{(1)},\tilde p_i^{(2)},\tilde p_j^{(1)},\tilde p_j^{(2)})$, $\tilde p_{i,j}^{(k)}=\tilde p_{i,j}^{k}(\hat\theta_1+t,\ldots,\hat\theta_N+t)$, and $Z^{(k)} = Z^{(k)}(\hat \theta_i+t)$. The differential equation above represents a system of non-autonomous ODEs for the phase dynamics of each oscillator.
		
		In order to obtain an autonomous ODE that preserves the long-term dynamics of this non-autonomous system, we apply (first-order) averaging theory to obtain:
		\begin{equation}\label{eq:th_avg}
			\dot{\theta}_i = \ve \sum_{j=1}^N a_{ij} \mathcal{H}^{(1)}(\theta_i,\theta_j) + \ve^2\sum_{j=1}^N a_{ij} \mathcal{H}^{(2)}(\theta_i,\theta_j) + \ve^3\sum_{j=1}^N a_{ij} \mathcal{H}^{(2)}(\theta_i,\theta_j)+\ldots,
		\end{equation}
		where
		\begin{align*}
			\mathcal{H}^{(1)}(\theta_i,\theta_j) &= \frac{1}{T} \int_0^T K^{(0)}\cdot Z^{(0)} dt,\\
			\mathcal{H}^{(2)}(\theta_i,\theta_j) &= \frac{1}{T} \int_0^T K^{(1)}\cdot Z^{(0)} + \tilde p_i^{(1)} K^{(0)}\cdot Z^{(1)}dt,\\
			\mathcal{H}^{(3)}(\theta_i,\theta_j) &= \frac{1}{T} \int_0^T K^{(2)}\cdot Z^{(0)} + \tilde p_i^{(1)}K^{(1)} \cdot Z^{(1)} + \tilde p_i^{(2)}K^{(0)} \cdot Z^{(1)} + \left(\tilde p_i^{(1)}\right)^2 K^{(0)}\cdot Z^{(2)} dt.
	\end{align*}}
	\yp{System \eqref{eq:th_avg} represents the phase dynamics of $N$ strongly coupled oscillators taking into account the amplitude dynamics.
		
		\textbf{Remark:} our use of first-order averaging is a strong assumption, but its utility depends on the system of interest. For the example systems we consider in this paper, first-order averaging is sufficient to capture phase dynamics far beyond the weak coupling regime. However, if a system or problem demands higher-order averaging, we may incorporate methods from \cite{llibre2014higher,maggia2020higher} in future studies}.
	
	We note that for numerical implementation, computing Equation \eqref{eq:p_i} is the most computationally expensive step because it a scalar time integral that must be recomputed for pairs of phase variables. We refer the reader to Appendix \ref{a:integral} for details of our numerical approach.
	
	\subsection{Computation of Coupling Functions for $N=2$ Oscillators}\label{sec:nonweak_example}
	
	As a concrete example of how our symbolic script generates phase equations, we show the process for deriving the phase equations for two reciprocally coupled oscillators $\theta_1$, $\theta_2$ up to order $O(\ve^3)$. We assume a system of $N=2$ coupled oscillators without self-coupling ($a_{ii}=0$). We write $\eta_i = \hat \theta_i+t$ for brevity.
	
	Recall the $\ve$-expansion in the coupling function $G$ in the previous section:
	\yp{
		\begin{equation*}
			\begin{split}
				G(\eta_1,\psi_1,\eta_2,\psi_2)= & K^{(0)}(\eta_1,\eta_2)\\
				&+ \ve K^{(1)}\left(\eta_1,\eta_2\yp{,p_1^{(1)},p_2^{(1)}}\right)\\
				&+ \ve^2 K^{(2)}\left(\eta_1,\eta_2,\yp{p_1^{(1)},p_1^{(2)},p_2^{(1)},p_2^{(2)}}\right)\\
				&+ \ldots.
			\end{split}
	\end{equation*}}
	Each $K^{(k)}$ contains the amplitude expansions $\psi_1$ and $\psi_2$. \yp{We derive more explicit forms} for $K^{(k)}$ by plugging in the Floquet eigenfunction expansions
	\begin{equation*}
		\Delta x_i \approx \psi_i(t) g^{(1)}(\eta_i) + \psi_i^2 g^{(2)}(\eta_i) + O\left(\psi_i^3\right),
	\end{equation*}
	into the derivative expansion of $G$,
	\begin{align*}
		G_m(Y + \Delta X)&= G_m(Y)+ G_m^{(1)}(Y)\Delta X+ \sum_{k=2}^\infty \frac{1}{k!}\left[ \stackrel{k}{\otimes} \Delta X^T\right] \vc\left(G_m^{(k)}(Y)\right),
	\end{align*}
	and collect in powers of $\psi_i$. The appropriately-collected terms are
	\begin{equation}\label{eq:G_psi}
		G(\eta_1,\yp{\psi_1},\eta_2,\yp{\psi_2}) = \sum_{k+\ell\leq 2} \psi_1^k \psi_2^\ell M^{(k,\ell)}(\eta_1,\eta_2),
	\end{equation}
	where the functions $M^{(k,\ell)}$ are maps $M^{(k,\ell)}:S^1\times S^1 \rightarrow \mathbb{R}^2$ consisting of the expanded Floquet eigenfunctions of order $\psi^k\psi^\ell$and the partial derivatives of $G$. To obtain an expansion in $\ve$, we plug in the amplitude expansion \eqref{eq:psi_exp} and collect in powers of $\ve$, noting that for $N=2$ without self coupling, the $p_i^{(k)}(t)$ terms are
	\begin{align*}
		p_i^{(k)}(t) &= \int_0^{\infty} e^{\kappa s}f^{(k)}(\hat\theta_i+t-s,\hat\theta_j+t-s)ds\\
		&\equiv p_i^{(k)}(\hat\theta_i+t,\hat\theta_j+t), 
	\end{align*}
	where $i=1,2, j=3-i$. The resulting $K^{(k)}$ functions are,
	\begin{align*}
		K^{(0)}(\eta_1,\eta_2) &= M^{(0,0)}(\eta_1,\eta_2),\\
		K^{(1)}(\eta_1,\eta_2) &= p_2^{(1)}(\eta_2,\eta_1)M^{(0,1)}(\eta_1,\eta_2) + p_1^{(1)}(\eta_1,\eta_2) M^{(1,0)}(\eta_1,\eta_2),\\
		K^{(2)}(\eta_1,\eta_2) &= \left(p_2^{(1)}(\eta_2,\eta_1)\right)^2 M^{(0,2)}(\eta_1,\eta_2) + p_1^{(1)}(\eta_1,\eta_2)p_2^{(1)}(\eta_2,\eta_1)M^{(1,1)}(\eta_1,\eta_2),\\
		&\quad\quad+ \left(p_1^{(1)}(\eta_1,\eta_2)\right)^2 M^{(2,0)}(\eta_1,\eta_2).
	\end{align*}
	Next, we write the $\ve$-expansion of the PRC function $\mathcal{Z}$, again using the amplitude expansion  \eqref{eq:psi_exp}:
	\begin{align*}\label{eq:z_expand}
		\mathcal{Z}(\eta_1,\ypb{\eta_2}) &= Z^{(0)}(\eta_1) + \ve p_1^{(1)}(\eta_1,\eta_2) Z^{(1)}(\eta_1) \\
		&\quad\quad+\ve^2\left[p_1^{(2)}(\eta_1,\eta_2)Z^{(1)}(\eta_1) + \left(p_1^{(1)}(\eta_1,\eta_2)\right)^2Z^{(2)}(\eta_1)\right]+O(\ve^3),
	\end{align*}
	where $i=1,2, j=3-i$. We now plug the $\ve$-expansions for $G$ and $\mathcal{Z}$ into the phase equation  \eqref{eq:th_moving}, and average to yield,
	\begin{equation*}
		\theta_1 = \ve \h^{(1)}(\theta_2-\theta_1) + \ve^2 \h^{(2)}(\theta_2-\theta_1) + \ve^3 \h^{(3)}(\theta_2-\theta_1) + O\left(\ve^4\right),
	\end{equation*}
	where
	\begin{align*}
		\h^{(1)}(\eta) &= \frac{1}{T} \int_0^T Z^{(0)}\cdot M^{(0,0)}ds.\\
		\h^{(2)}(\eta) &= \frac{1}{T} \int_0^T p_1^{(1)}Z^{(1)}\cdot M^{(0,0)} + p_2^{(1)} Z^{(0)}\cdot M^{(0,1)} + p_1^{(1)} Z^{(0)} M^{(1,0)} ds.\\
		\h^{(3)}(\eta) &= \frac{1}{T} \int_0^T \yp{\left[ Z^{(0)}\cdot K^{(2)}  +p_1^{(1)} Z^{(1)}\cdot K^{(1)}+ p_1^{(2)}Z^{(1)}\cdot K^{(0)} + \left(p_1^{(1)}\right)^2Z^{(2)}\cdot K^{(0)} \right] ds.}
	\end{align*}
	All $Z^{(k)}$ are functions of the integrating variable $s$, all \yp{$K^{(k)}$} are functions of $(s,\eta+s)$, all $p_1^{(k)}$ are functions of $(s,\eta+s)$, and all $p_2^{(k)}$ \yp{(contained inside the $K^{(k)}$ terms)} are functions of $(\eta+s,s)$. The equation for $\theta_2$ is identical, but with $\theta_1-\theta_2$ as inputs to the $\h^{(k)}$ functions.
	
	Finally, we take the phase difference $\phi = \theta_2 - \theta_1$, resulting in the scalar equation,
	\begin{equation*}
		\begin{split}
			\dot{\phi} &= \ve \left[\h(-\phi) - \h(\phi) \right]\equiv -2\h_\text{odd}(\phi)\\
			&=\ve \left[\h^{(1)}(-\phi) - \h^{(1)}(\phi) \right] + \ve^2 \left[\h^{(2)}(-\phi) - \h^{(2)}(\phi) \right]\\
			&\quad + \ve^3 \left[\h^{(3)}(-\phi) - \h^{(3)}(\phi) \right] + O\left(\ve^4\right),
		\end{split}
	\end{equation*}
	where $\h_\text{odd}$ is the odd part of $\h$, and we will often refer to the right-hand side of the above as $-2\h_\text{odd}$, or with a slight abuse of notation, simply call them interaction functions. As mentioned earlier, fixed points of the scalar $-2\h_\text{odd}$ function inform us of the existence and stability of phase-locked states in a given pair of coupled oscillators. We will demonstrate this property in the examples to follow.
	
	We remark that this expansion is consistent with previously developed strategies for getting the first and second order responses. The first order term contains the functions $M^{(0,0)}$ and $Z^{(0)}$, which are the coupling function $G$ and classic infinitesimal phase response curve, so $\h^{(1)}$ is the classic interaction function. The second-order term is identical to that derived in \cite{wilson2019phase}. Most importantly, the proposed theory represents an extension of the work in \cite{wilson2019phase} and can be used to calculate coupling functions to arbitrary orders of accuracy in the coupling strength.  Higher order approximations are straightforward to attain through symbolic manipulations, which is automated using our Python code.

	\section{Results}
	\subsection{CGL Model}\label{sec:cgl}
	
	\begin{figure}[ht!]
		\centering
		\includegraphics[width=\textwidth]{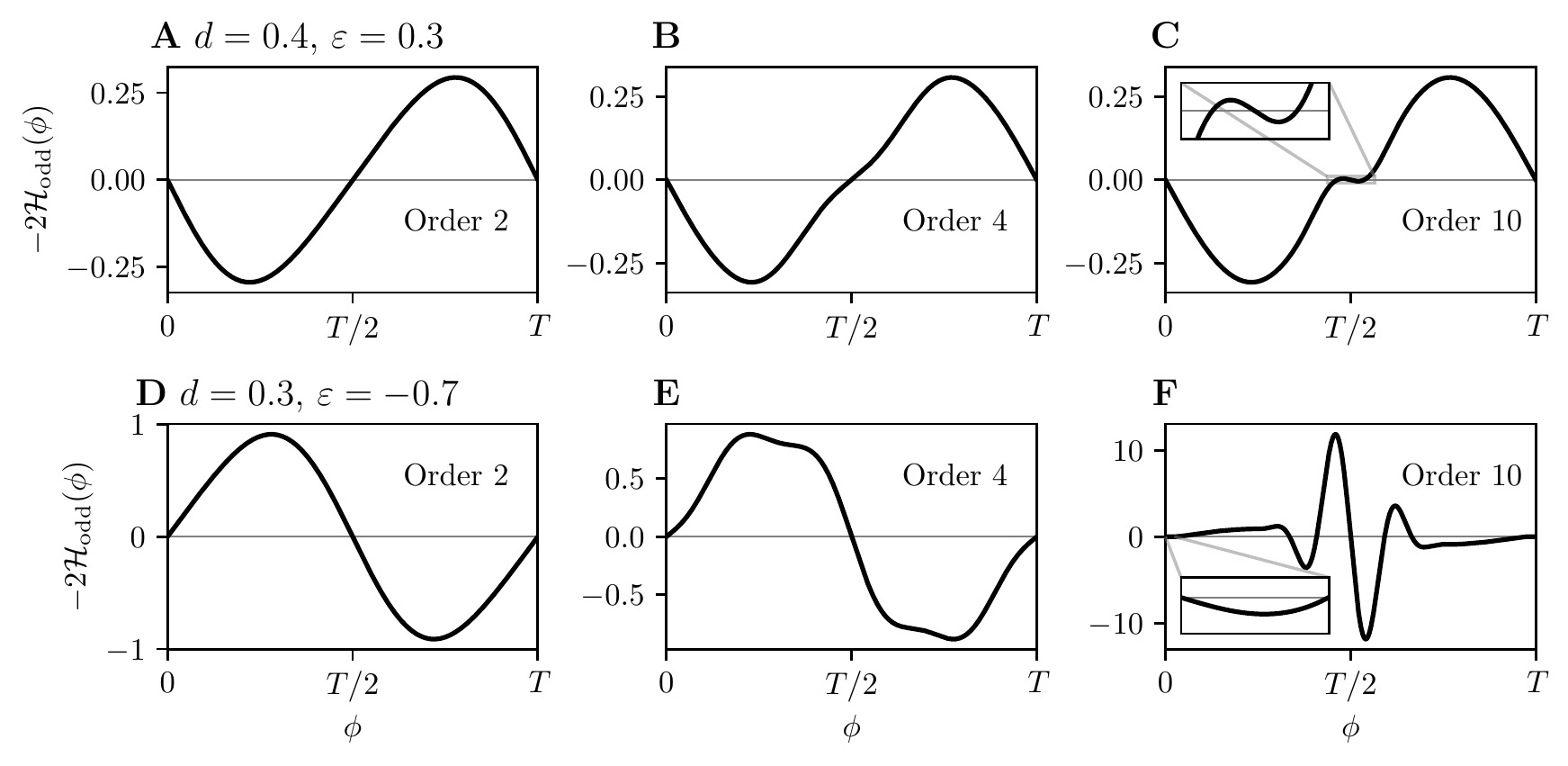}
		\caption{Examples of generalized $\h$ functions in the CGL model. Roots indicate existence of phase-locked solutions (with stability determined by the slopes). A,B,C: second, fourth, and tenth order interaction functions, respectively, for the choice of coupling parameters $d=0.4$ and $\ve=0.26$. Stability in the antiphase state only appears in C with the addition of the tenth-order term (inset shows a negative slope at antiphase). D,E,F: higher-order coupling functions for second, fourth, and tenth order, respectively, for the choice of coupling parameters $d=0.3$ and $\ve=-0.66$. Stability in the synchronous state only appears in F with the addition of the tenth-order term (inset shows a negative slope at synchrony). $q=1$ for this example.}\label{fig:cgl_h}
	\end{figure}
	
\begin{figure}[ht!]
		\centering
		\includegraphics[width=.75\textwidth]{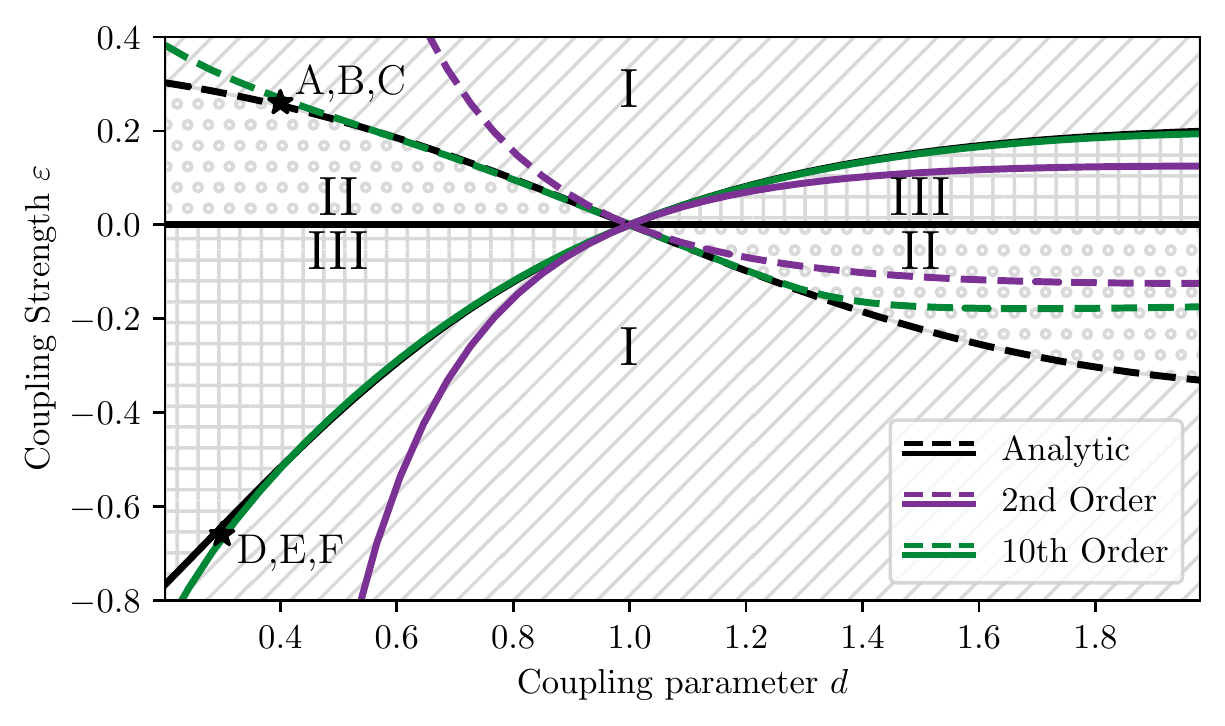}
		\caption{Two-parameter diagram of the CGL model in coupling parameters $d$ and $\ve$. Synchrony is only stable in regions I and II, whereas antiphase is only stable in regions I and III. All black lines are analytically computed from Equation \eqref{eq:cgl_true}. Black solid lines denote boundaries where the system switches between stable and unstable synchrony ($\ve_s$). Black dashed lines denote boundaries where the system switches between stable and unstable antiphase ($\ve_a$). Purple solid, dashed: bifurcations detected using 2nd order interaction functions from \cite{wilson2019phase}. Green solid, dashed: bifurcations detected using 10th order interaction functions. The points labeled $\star$A,B,C and $\star$D,E,F correspond to the parameter values used in Figure \ref{fig:cgl_h}A,B,C, and D,E,F, respectively.}\label{fig:cgl_2par}
	\end{figure}

	We begin with a relatively straightforward example of two diffusively coupled complex Ginzburg-Landau (CGL) models:
	\begin{align}\label{cgl}
		x_j' &= (1-x_j^2-y_j^2)x_j - q(x_j^2+y_j^2)y_j + \ve \left[x_k - x_j - d(y_k-y_j)\right], \nonumber \\
		y_j' &= (1-x_j^2-y_j^2)y_j + q(x_j^2+y_j^2)x_j + \ve \left[ y_k - y_j + d(x_k-x_j)\right],
	\end{align}
	where $j=3-k$ with $k=1,2$. When $\ve=0$ and $q=1$, the system admits a stable $2\pi$-periodic limit cycle, $x_j(t) = \cos(qt)$, $y_j=\sin(qt)$.  Depending on the choices of $d$, $q$ and $\epsilon$, the model \eqref{cgl} can admit stable phase locked solutions, stable antiphase solutions, or bistability between phase-locked and antiphase solutions.  Critical curves that define regions of stability were computed exactly by  \cite{aronson1990amplitude} and are given by
	\begin{equation}\label{eq:cgl_true}
		\begin{split}
			\ve_s &= \frac{dq-1}{d^2+1},\\
			\ve_a &= \frac{1-dq}{d^2-2dq+3}.
		\end{split}
	\end{equation}
	These curves are shown as black lines in Figure \ref{fig:cgl_2par} and define regions where different locking modalities are stable. We compare our method to the ground truth of Equation \eqref{eq:cgl_true} by generating $\h$ functions of different order truncations and tracking the fixed points of the equation
	\begin{align*}
		\dot\phi &= \ve\left[\h(-\phi) - \h(\phi)\right].
	\end{align*}
	as a function of $\ve$ and $d$. 
	
	Examples of $\h$ functions are shown in Figure \ref{fig:cgl_h}. Panels A, B, and C show the $\h$ function for second, fourth, and tenth order for $d=0.4$ and $\ve=0.26$. For this coupling strength, second and fourth order $\h$ functions show that antiphase is unstable, but the tenth order function reveals a stable antiphase solution. Panels D, E, F, show the $\h$ function for second, fourth, and tenth order for $d=0.3$ and $\ve=-0.66$.  Second and fourth order show that synchrony is unstable, but the tenth order function reveals a stable synchronous solution. 
	
	In Figure \ref{fig:cgl_2par}, we show the boundaries constructed from our theory using second order (purple) and tenth order (green) $\h$ functions. The system switches between stable and unstable synchrony across solid lines (between regions I and III), and between stable and unstable antiphase across dashed lines (between regions I and II). As expected, the tenth-order approximation closely follows the ground-truth curves (black) for a much greater range of $d,\ve$, compared to existing second-order theory. The parameter values corresponding to Figure \ref{fig:cgl_h}A, B, C are labeled with a star ($\star$) towards the upper left corner of the diagram. This point is in parameter region I, which corresponds to stable synchrony and unstable antiphase, confirming the accuracy of Figure \ref{fig:cgl_h}C. The parameter values corresponding to Figure \ref{fig:cgl_h}D, E, F are labeled with a star ($\star$) towards the lower left corner of the diagram. This point is also in parameter region I and we confirm stable synchrony and unstable antiphase observed in Figure \ref{fig:cgl_h}F.
	
	The analytically tractable features of this model allow us to confirm our theory and demonstrate its strong performance. Additionally, our theory can also be applied straightforwardly to analytically intractable models as will be seen in the next example.
	
	\subsection{Thalamic Neuron Model}\label{sec:thal}
	
	As a second example, we consider a model of synaptically coupled conductance-based neurons taken from \cite{rubi04} that replicate the salient dynamical behaviors of tonically firing thalamic neurons. The state variables of the thalamic neuron models satisfy
	\begin{align*}
		\frac{dV_i}{dt} &= (-I_\text{L}(V_i) + I_{\text{Na}}(V_i) + I_\text{K}(V_i) + I_\text{T}(V_i) - g_\text{syn}w_j(V_i-E_\text{syn})+I_\text{app})/C,\\
		\frac{dh_i}{dt} &= (h_\infty(V_i) - h_i)/\tau_h(V_i),\\
		\frac{dr_i}{dt} &= (r_\infty(V_i) - r_i)/\tau_r(V_i),\\
		\frac{dw_i}{dt} &= \alpha(1-w_i)/(1+\exp((V_i-V_\text{T})/\sigma_T)) - \beta w_i,
	\end{align*}
	
	where $i=1,2$, $j=3-i$. The voltage variable $V_i$ depend on the gating variables $h_i, r_i$, and receives synaptic inputs from the synaptic variable $w_j$ from the reciprocal neuron. We consider excitatory synaptic coupling, $E_\text{syn} = 0$. We will use the parameter $g_\text{syn}$ to denote the coupling strength in this section (it is equivalent to $\ve$ in our formulation). All remaining equations are listed in Appendix \ref{a:thal} along with the parameters in Table \ref{tab:thal}. 
	
	\begin{figure}[ht!]
		\centering
		\includegraphics[width=\textwidth]{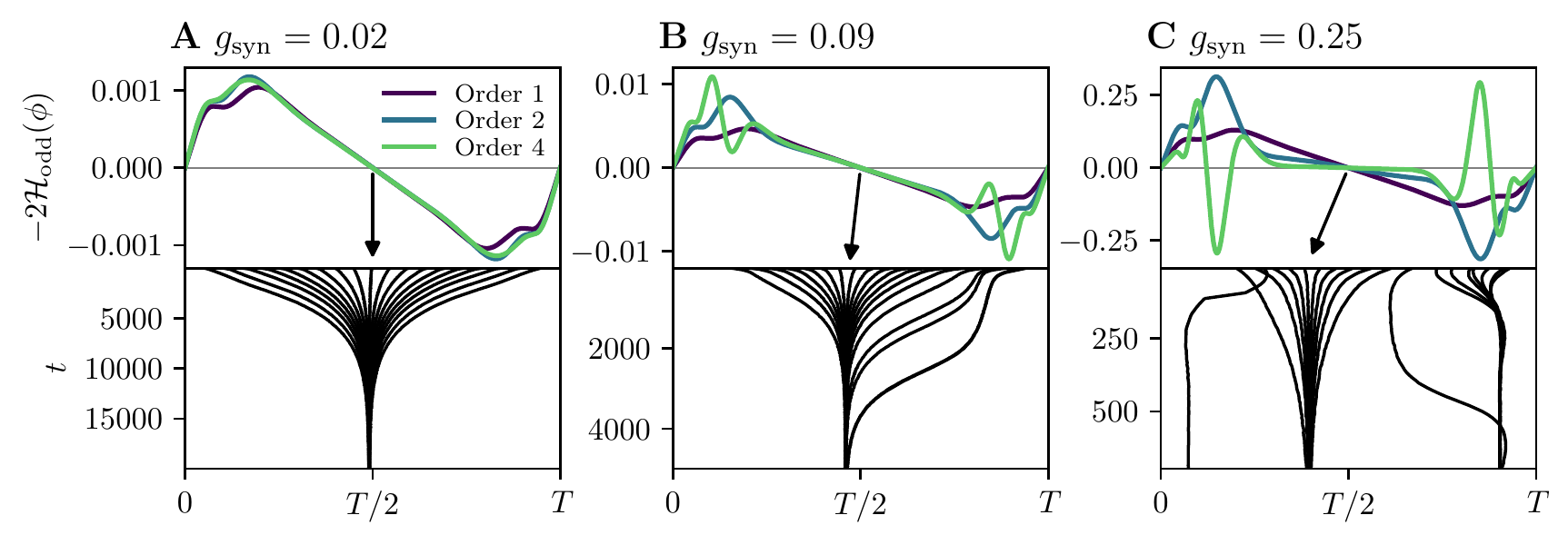}
		\caption{Examples of generalized $\h$ functions in the thalamic model. Roots indicate existence of phase-locked solutions, and slopes the stability. In each top panel, first (purple), second (blue), and fourth (green) order generalized interaction functions are shown. In each bottom panel, the phase difference between two \yp{full} thalamic models are shown for 20 initial conditions. A: with $g_\text{syn}=0.02$, i.e., weak coupling, all orders agree and the full model converges to the antiphase state as indicated by the black arrow. B: with $g_\text{syn}=0.0.9$, the weak coupling theory remains valid and the stability of fixed points agree with the higher-order interaction functions. \yp{Note that the fourth order function (top panel) predicts a region indicated by $\star$, where phase differences evolve relatively slowly. The phase differences near synchrony in the full model (bottom panel) exhibit slow changes in the phase differences indicated by a $\star$, consistent with the fourth-order prediction}. C: with $g_\text{syn}=0.25$, only fourth order captures the existence of near-synchronous states. To ease comparisons, we scaled the first order function by a factor of 10 and the second order term by a factor of 7. The black arrow indicates the location of the antiphase point in the full system.}\label{fig:thal_h}
	\end{figure}
	
	Figure \ref{fig:thal_h} shows generalized $\h$ functions up to first, second, and fourth order for $g_\text{syn}\in \{0.01, 0.09, 0.25\}$. For $g_\text{syn}=0.02$, all generalized $\h$ functions exhibit the same types of stability, namely unstable synchrony and stable antiphase (Figure \ref{fig:thal_h}A), and the full model converges to antiphase as expected (Figure \ref{fig:thal_h}A, black arrow and black curves). For $g_\text{syn}=0.09$, all $\h$ functions agree in stability (Figure \ref{fig:thal_h}B, bottom black arrow and black lines). However, only the fourth-order $\h$ function explains the slow transitions to antiphase for solutions near synchrony.
	
	We remark on a few important features in the bottom panels of Figure \ref{fig:thal_h}B,C that may appear erroneous but are in fact consistent with our theory. Note that the underlying limit cycle will deform as a function of the coupling strength $g_\text{syn}$, and the greater the coupling strength the greater the deformation. Although we don't show the limit cycle deformation explicitly, we have observed that the shape and period of the limit cycle with no coupling, $g_\text{syn}=0$, may differ substantially from the shape and period of the limit cycle for stronger coupling, e.g., when we increase $g_\text{syn}$ to $g_\text{syn}=0.09$ and $g_\text{syn}=0.25$. In the case of weak coupling, a standard approach is to use the limit cycle with no coupling as a reference point to initialize solutions with some desired phase difference. This choice works well because weak coupling does not perturb solutions far from the unperturbed limit cycle. However, in the case of strong coupling, using the unperturbed limit cycle to initialize solutions results in strong transients as the solutions settle on to the strongly perturbed limit cycle. Because the strongly perturbed limit cycle may differ greatly in shape and period from the unperturbed limit cycle, the transients sometimes allow oscillators to switch in the sign of the phase.
	
	For example, oscillators initialized using the unperturbed limit cycle at $\theta_1$ and $\theta_2$, where $\theta_2$ lags just behind $\theta_1$ ($\phi = \theta_2-\theta_1 < 0$), may rapidly switch in order and result in $\theta_1$ lagging just behind $\theta_2$ ($\phi > 0$) as the underlying trajectories settle on to the strongly perturbed limit cycle. It is possible to mitigate the issue of transients by using the strongly perturbed limit cycle as a reference to initialize solutions, but we chose to use the unperturbed limit cycle as it is a standard approach. For the few initial conditions that result in this type of switch, we chose to reverse their sign \textit{post hoc}. For this reason, some initial conditions appear to be missing in Panel B, and some phase difference trajectories overlap in panel C.
	
	Regarding the convergence of phase differences away from the antiphase $T_0/2$ in panels B and C, note that we used the period $T_0\approx 10.6$ of the unperturbed oscillator to normalize all solutions, so the antiphase state during strong coupling will appear incorrect by a factor of $T_{0.09}/T_0$ and $T_{0.25}/T_0$, where $T_{0.09}\approx 10$ is the period of oscillation at $g_\text{syn}=0.09$ and $T_{0.25}\approx 8.4$ is the period of oscillation at $g_\text{syn}=0.25$. The ratios $T_{0.09}/T_0$ and $T_{0.25}/T_0$ are consistent with the respective differences seen in panels B and C.
	
	\begin{figure}[ht!]
		\centering
		\includegraphics[width=\textwidth]{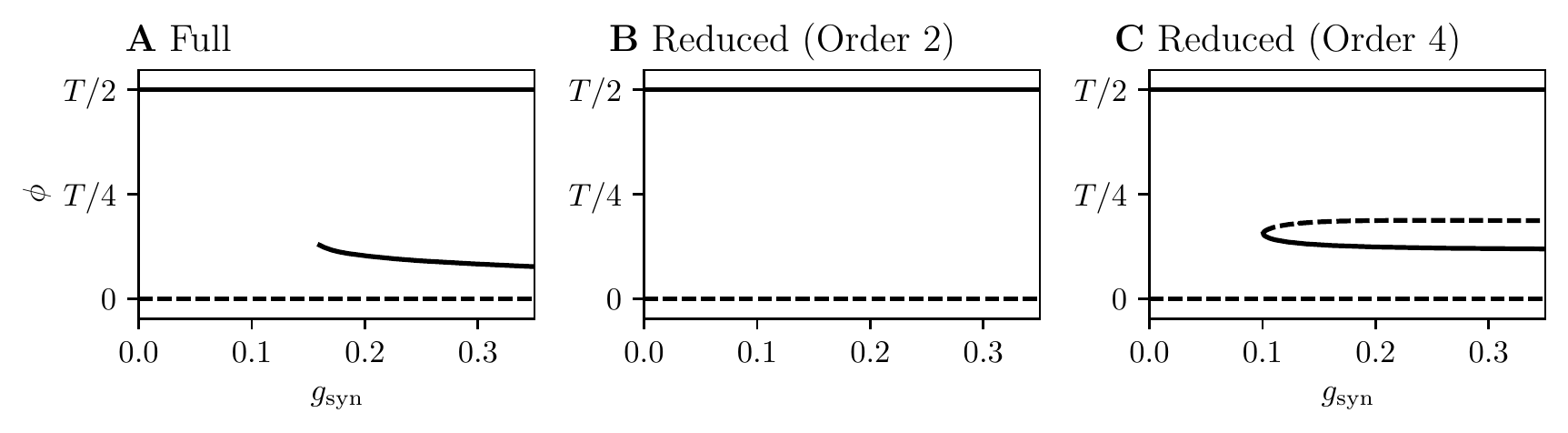}
		\caption{One-parameter bifurcation diagrams in $g_\text{syn}$ of the phase difference between two Thalamic oscillators. A: Bifurcation diagram of the full system. Synchrony is unstable, antiphase is stable, and for \yp{$g_\text{syn}\approx 0.19$}, a stable near-synchronous state emerges. B: Bifurcation diagram of the reduced system using an order 2 approximation.   Synchrony is unstable and antiphase is stable as expected, but there is no near-synchronous solution.  The bifurcation diagram when using the order 1 accurate coupling functions (i.e.~the standard theory of weakly coupled oscillators) is identical to the order 2 accurate diagram.  C: Bifurcation diagram of the reduced system using an order 4 approximation. Synchrony is unstable and antiphase is unstable in agreement with the full model, and the near-synchronous branch appears for $g_\text{syn}\approx 0.1$.}\label{fig:thal_1par}
	\end{figure}
	
	We further illustrate the differences between the generalized $\h$ functions using one-parameter bifurcation diagrams (Figure \ref{fig:thal_1par}). Similar to the CGL model, we follow fixed points of the phase difference equation
	\begin{align*}
		\dot\phi &= g_\text{syn}\left[\h(-\phi) - \h(\phi)\right],
	\end{align*}
	for different order truncations. The bifurcation parameter is naturally $g_\text{syn}$.
	
	\yp{To compute the one-parameter diagram of the full model, we used Newton's method to converge onto the underlying stable phase-locked states. For a given coupling strength $g_\text{syn}$, we initialized the model at antiphase to capture the antiphase solution, then incremented $g_\text{syn}$ by a small amount ($0.02$nS) and repeated the stability calculation to follow the antiphase branch. To compute the stability of other branches, we repeated this calculation by initializing the full model at different phase shifts (one at synchrony and the other at a phase difference that led to the stable phase-locked branch).} The full model exhibits unstable synchrony, stable antiphase, and a stable phase-locked state that emerges at \yp{$g_\text{syn} \approx 0.19$} as $g_\text{syn}$ increases (Figure \ref{fig:thal_1par}A). \yp{We were unable to capture unstable phase-locked branches in the full system using this method}.
	
	In Figure \ref{fig:thal_1par}, we find that using the second-order $\h$ function captures unstable synchrony and stable antiphase, but not the stable phase-locked solution. Finally, using the fourth-order $\h$ function, we capture all qualitative features of the full model including the stable phase-locked solution, which emerges at $g_\text{syn} \approx 0.1$. We are also able to capture the unstable phase-locked branch.
	
	This result demonstrates the general utility our theory. It is naturally applicable to arbitrary, smooth $n$-dimensional smooth dynamical systems with arbitrary, smooth coupling functions. Despite stronger coupling inducing relatively large changes to the underlying vector field, the theory robustly reproduces the behaviors of the full, unreduced model.
	
	\section{Discussion}\label{sec:discussion}
	
	In this paper, we have established a general coupled oscillator theory for coupling strengths that extend well beyond the regime of weak coupling. By exploiting phase-amplitude relationships based on the isostable coordinate framework, we derived coupling functions valid to arbitrary orders of accuracy in the coupling strength. To verify the theory, we applied our theory to two different models. In the first example, we used the CGL model to demonstrate how higher-order coupling functions accurately characterize both the existence and stability of synchronous and antiphase solutions. Using these higher-order coupling functions, we reproduced the analytically derived boundaries in the two-parameter bifurcation diagram with much greater accuracy than existing methods. In the second example, we considered a neurobiologically motivated model of a tonically firing neuron. Our theory accurately reproduced the phase-locked solutions of coupled thalamic models.
	
	Provided relatively mild conditions such as sufficient smoothness of the vector fields are satisfied, our theory can be applied to a wide variety of oscillatory dynamical systems in the biological, chemical, and physical sciences. While we only explicitly considered $N=2$ oscillators in this paper, an important future direction includes augmenting this theory to networks of oscillators and extending classic results on weakly coupled oscillators.
	
	We have demonstrated the utility of our theory, but some limitations remain. First, strong coupling leads to strongly deformed limit cycles in the full system, and phase information from the weakly coupled system does not necessarily transfer into the strongly coupled system. While our theory \yp{manages to accurately capture phase and amplitude information far from the unperturbed limit cycle without using direct} knowledge of the strongly coupled system, care must be taken when translating from our theory to the full system. The theory is best suited to understanding the \textit{asymptotic} behavior of coupled oscillators (although it is worth reiterating that the theory can reproduce qualitative transient behavior for non-weak coupling).
	
	Other limitations of our theory are computational. Some of the functions in this paper are expensive to compute, but this limitation may be improved by existing work on phase reduction theory for strong perturbations. In \cite{letson2020lor}, the authors introduce the local orthogonal rectification (LOR). In contrast to the isostable framework, LOR codes the amplitude as an orthogonal distance from a limit-cycle trajectory.  Other insights that may lead to more efficient computation of coupling functions may be gleaned from \cite{perez2020global}, where authors introduce a parameterization method to compute higher-order phase-amplitude coordinates, which sidesteps the need to compute symbolic derivatives and the need to use Newton's method in this work and in \cite{wilson2019phase}.
	
	\yp{Finally, we discuss where our results stand relative to general work using pulse-coupled oscillators. In \cite{canavier2009phase}, the authors derive a general method -- independent of model, coupling strength, and synapse -- to predict $N:1$ phase locking. These pulse-coupled methods are powerful and broadly applicable to experimental neuroscience because the underlying differential equations need not be known. Similarly, Cui et al. derive a functional phase response curve given a regular stream of pulse trains perturbing oscillator phase responses with additional effects such as adaptation\cite{cui2009functional}. However, the former results rely on a strongly attracting limit cycle and the latter results on the input type (pulsatile and regular). Beyond the linear regime, if the input changes or multiple inputs are applied, it is generally challenging to generalize the experimentally obtained phase response curves. In particular, when only considering the linear phase response curves, the resulting reduced order equations, in general (especially with weakly attracting limit cycles), will not predict bifurcations that result as the coupling strength increases. The continuous-time method proposed in this paper provides a systematic method for generating coupling functions valid to higher than linear orders of accuracy in the reduced order coordinates.}
	
	\appendix
	
	\setcounter{table}{0}
	\renewcommand{\thetable}{A\arabic{table}}
	\yp{
		\section{Derivation of the Taylor Expansion of $G$ in $\ve$}\label{a:K}
		
		Recall that starting with an arbitrary $m$th coordinate of the vector-valued function $G(Y+\Delta X)$, where $Y=[Y(\theta_i)^T,Y(\theta_j)^T]^T$ and $\Delta X = [\Delta x_i^T, \Delta x_j^T]^T$ \yp{(both $2n \times 1$ column vectors)}, the Taylor expansion yields Equations \eqref{eq:g_raw} and \eqref{eq:G0}, rewritten here for convenience:
		\begin{align}\label{eq:g_raw_a}
			G_m(Y + \Delta X)&= G_m(Y)+ G_m^{(1)}(Y)\Delta X+ \sum_{k=2}^\infty \frac{1}{k!}\left[ \stackrel{k}{\otimes} \Delta X^T\right] \vc\left(G_m^{(k)}(Y)\right),
		\end{align}
		where 
		\begin{equation}\label{eq:G0_a}
			G_m^{(k)} = \frac{\pa \vc\left(G_m^{(k-1)}\right)}{\pa X^T} \in \mathbb{R}^{2n^{(k-1)}\times 2n}.
		\end{equation}
		The goal is to arrive at the form in Equation \eqref{eq:G_eps}:
		\begin{equation}\label{eq:G_eps_a}
			G(\theta_i,\psi_i,\theta_j,\psi_j)= K^{(0)}(\theta_i,\psi_i,\theta_j,\psi_j) + \ve K^{(1)}(\theta_i,\psi_i,\theta_j,\psi_j) + \ve^2 K^{(2)}(\theta_i,\psi_i,\theta_j,\psi_j) + \ldots
		\end{equation}
		
		To begin, we substitute the Floquet eigenfunction for $\Delta x_i$,
		\begin{equation}\label{eq:delta_xi_a}
			\Delta x_i = \psi_i g^{(1)}(\theta_i) + \psi_i^22 g^{(2)}(\theta_i) +\ldots,
		\end{equation}
		into Equation \eqref{eq:g_raw_a}. For clarity, we will consider individual terms in Equation \eqref{eq:g_raw_a} and later collect in powers of $\ve$. The first term, $G_m(Y)$, is a scalar and is $O(1)$ in $\ve$ This term is equivalent to $K^{(0)}$. The second term contains $G_m^{(1)}(Y)$, a $1\times 2n$ vector that multiplies $\Delta X$, a $2n\times 1$ vector, therefore this term is a scalar, as expected. Written in terms of Equation \eqref{eq:delta_xi_a},
		\begin{equation}\label{eq:G_m2}
			G_m^{(1)}(Y) \Delta X = G_m^{(1)}(Y)
			\left[ \begin{matrix}\psi_i g^{(1)}(\theta_i) + \psi_i^2 g^{(2)}(\theta_i)+\ldots\\
				\psi_j g^{(1)}(\theta_j) + \psi_j^2 g^{(2)}(\theta_j)+\ldots\end{matrix}\right],
		\end{equation}
		where each $g^{(k)}$ is a $2n \times 1$ vector. If we then substitute the expansion for $\psi_i$,
		\begin{equation*}\label{eq:psi_exp_a}
			\psi_i(t) \approx \ve p_i^{(1)}(t) + \ve^2 p_i^{(2)}(t) + \ve^3 p_i^{(3)}(t) + \ldots,
		\end{equation*}
		into Equation \eqref{eq:G_m2}, we arrive at a form where powers of $\ve$ are explicit:
		\begin{equation}\label{eq:g_m_a}
			\begin{split}
				&G_m^{(1)}(Y) \Delta X\\
				&= G_m^{(1)}(Y)
				\left[ \begin{matrix}
					\left(\ve p_i^{(1)}(t) + \ve^2 p_i^{(2)}(t) + \ldots\right) g^{(1)}(\theta_i) + \left(\ve p_i^{(1)}(t) + \ve^2 p_i^{(2)}(t) + \ldots\right)^2 g^{(2)}(\theta_i)+\ldots\\
					\left(\ve p_j^{(1)}(t) + \ve^2 p_j^{(2)}(t) + \ldots\right) g^{(1)}(\theta_j) + \left(\ve p_j^{(1)}(t) + \ve^2 p_j^{(2)}(t) + \ldots\right)^2 g^{(2)}(\theta_j)+\ldots
				\end{matrix}\right]\\
				&= G_m^{(1)}(Y)
				\left(\ve\left[\begin{matrix}
					p_i^{(1)}(t) g^{(1)}(\theta_i)\\
					p_j^{(1)}(t) g^{(1)}(\theta_j)
				\end{matrix}\right]
				+ \ve^2\left[\begin{matrix}
					p_i^{(2)}(t)g^{(1)}(\theta_i) + p_i^{(1)}(t)^2 g^{(2)}(\theta_i)\\
					p_j^{(2)}(t)g^{(1)}(\theta_j) + p_j^{(1)}(t)^2 g^{(2)}(\theta_j)
				\end{matrix}\right] + \ldots\right).
			\end{split}
		\end{equation}
		Although not explicitly written here, it is straightforward to derive expressions for the higher-order terms in $\ve$, especially using a symbolic \ypb{math toolbox such as those found in Python, MATLAB/Octave, Mathematica, and Maple (we opted to use Python's Sympy \cite{sympy})}. Note that the first-order $\ve$ term above is equivalent to $K^{(1)}$ and this term only depends on up to $p_i^{(1)}$.
		
		We now turn to the next term in the Taylor expansion of $G$ in Equation \eqref{eq:g_raw_a}. This term contains a tensor product $\Delta X^T \otimes\Delta X^T $, which yields a $1 \times (2n)^2$ vector, and the $\vc(\cdot)$ operator applied to the second derivative of $G_m$, $\vc\left(G_m^{(2)}(Y)\right)$, which yields a $(2n)^2 \times 1$ vector. Therefore, the second term is a scalar as expected. All powers of $\ve$ are contained in the first term, so we unpack this term explicitly. 
		
		Let $g_m^{(k)}$ denote the $m$th coordinate of the vector-valued function $g^{(k)}$ and $\ypb{X}_m$ denote the $m$th coordinate of $\Delta X$ for $m=1,\ldots,2n$. Then
		\begin{align*}
			\ypb{X}_m = \left\{ \begin{matrix}
				\psi_i g_m^{(1)}(\theta_i) + \psi_i^2 g_m^{(2)}(\theta_i)+\ldots,&\text{for}\quad m=1,\ldots,n,\\
				\psi_j g_m^{(1)}(\theta_j) + \psi_j^2 g_m^{(2)}(\theta_j) +\ldots,&\text{for}\quad m=n+1,\ldots,2n. \end{matrix} \right.
		\end{align*}
		The tensor product $\Delta X^T \otimes\Delta X^T$ can then be written,
		\begin{align*}
			&\Delta X^T \otimes\Delta X^T\\
			&= \left[\ypb{X}_1^2, \ypb{X}_1\ypb{X}_2,\ldots, \ypb{X}_{2n}^2\right]\\
			&= \left[\left(\psi_i g^{(1)}_1(\theta_i)+\psi_i^2 g^{(2)}_1(\theta_i)+\ldots\right)^2,\right.\\
			&\quad \quad \left(\psi_i g^{(1)}_1(\theta_i)+\psi_i^2 g^{(2)}_1(\theta_i)+\ldots\right)\left(\psi_i g^{(1)}_2(\theta_i)+\psi_i^2 g^{(2)}_2(\theta_i)+\ldots\right), \ldots,\\
			&\left.\quad\quad \left(\psi_j g^{(1)}_{2n}(\theta_j)+\psi_j^2 g^{(2)}_{2n}(\theta_j)+\ldots\right)^2\right]\\
			&= \left[\left(\left\{\ve p_i^{(1)}(t) + \ve^2 p_i^{(2)}(t)+\ldots\right\} g^{(1)}_1(\theta_i)+\left\{\ve p_i^{(1)}(t)+ \ve^2 p_i^{(2)}(t)+\ldots\right\}^2 g^{(2)}_1(\theta_i)+\ldots\right)^2,\right.\\
			&\quad \quad \left(\left\{\ve p_i^{(1)}(t)+ \ve^2 p_i^{(2)}(t) +\ldots\right\} g^{(1)}_1(\theta_i)+\left\{\ve p_i^{(1)}(t) + \ve^2 p_i^{(2)}(t)+\ldots\right\}^2 g^{(2)}_1(\theta_i)+\ldots\right)\\
			&\quad\quad\quad\times \left(\left\{\ve p_i^{(1)}(t)+ \ve^2 p_i^{(2)}(t) +\ldots\right\} g^{(1)}_2(\theta_i)+\left\{\ve p_i^{(1)}(t)+ \ve^2 p_i^{(2)}(t)+\ldots\right\}^2 g^{(2)}_2(\theta_i)+\ldots\right), \ldots,\\
			&\left.\quad\quad \left(\left\{\ve p_j^{(1)}(t)+ \ve^2 p_j^{(2)}(t)+\ldots\right\} g^{(1)}_{2n}(\theta_j)+\left\{\ve p_j^{(1)}(t)+ \ve^2 p_j^{(2)}(t)+\ldots\right\}^2 g^{(2)}_{2n}(\theta_j)+\ldots\right)^2\right].
		\end{align*}
		It is now possible to collect in powers of $\ve^2$. The order $\ve^2$ terms above, combined with those in Equation \eqref{eq:g_m_a} belong to the term $K^{(2)}$. Note that the $K^{(2)}$ term's dependence on $\psi_i$ and $\psi_j$ only appears in the functions $p_i^{(1)}, p_i^{(2)}, p_j^{(1)}$, and $p_j^{(2)}$.
		
		In general, using a symbolic \ypb{math toolbox}, it is possible to verify up to the desired order that the $K^{(k)}$ term only depends on terms $p_i^{(\ell)}$, $p_j^{(\ell)}$ for $\ell \leq k$.
	}
	
	\section{Numerical Integration}\label{a:integral}
	
	For $N=2$, our theory requires the computation of the functions
	\begin{align*}
		p_i^{(k)}(\eta_1,\eta_2) &= \int_0^{\infty} e^{\kappa s}f^{(k)}(\eta_1-s,\eta_2-s)ds.
	\end{align*}
	The above integral must be repeated for each $\eta_i$ and $\eta_j$, where $\eta_i$ and $\eta_j$ are taken from a grid of phase values. If $M$ is the number of discretized points in the interval $[0,T]$, $P$ is the number of discretized points in the interval $(\infty,0]$, and $N$ is the number of oscillators, then the total number of computations is proportional to $P\times M^{N}$. This computation is especially expensive if $\kappa$ is small (requiring large $P$), or if the functions $f^{(k)}$, containing the underlying \yp{phase response curves (PRCs), isostable response curves (IPRCs)}, and Floquet eigenfunctions require a fine temporal resolution to integrate (requiring $M$ large). In the case of the thalamic model, $\kappa$ is small, roughly $\kappa\approx 0.023$, and at least 20000 time units were required to compute the PRC, IRC, and Floquet eigenfunctions to acceptable accuracy (in contrast, the CGL model required fewer time units on the order of 2000). \yp{We typically iterated Newton's method until the magnitude of the derivative vector reduced to $1e-10$ or lower, which resulted in the magnitude of the difference between the final and initial conditions of the periodic solutions to be on the order of $1e-7$}. We used $P=25\times M$, so each integral calculation was relatively costly. We found that the $\eta_i$ and $\eta_j$ was best sampled using $M=4000$, i.e., a grid of $4000\times 4000$ discretized phase values, so the integral was computed 16 million times.  We found Riemann integration to be efficient and sufficiently accurate \yp{(we chose the integration mesh such that further refinements did not visually affect the $\mathcal{H}$-functions)}.
	
	To speed up computations of the above integral, we transformed it to minimize repeating calculations in two variables. Letting $u=\eta_1-s$, a straightforward transformation yields
	\begin{equation*}
		e^{\kappa \eta_1}\left[\int_{-\infty}^{0} e^{-\kappa u} f(u,\eta_2-\eta_1 + u) du + \int_{0}^{\eta_1} e^{-\kappa u} f(u,\eta_2-\eta_1 + u) du \right].
	\end{equation*}
	Note that the first integral depends only on the phase difference $\eta_2-\eta_1$, so it is computed along only one dimension, and the total number of computations is proportional to $P\times M\times(N-1)$. The second integral does not solely depend on the phase difference $\eta_2-\eta_1$, and must be computed in two dimensions. However, the computation is not on the entire grid of $\eta_1,\eta_2$ points, but on a triangular half of the domain because the upper integral limit varies as a function of $\eta_1$. The number of computations for the second integral is proportional to $M^{N}/2$. The total number of calculations $P\times M\times(N-1)+M^{N}/2$, which is a significant reduction compared to the original integral.
	
	Finally, we vectorized and computed each integral independently for any given $\eta_1,\eta_2$, allowing us to parallelize the integral computation. The parallelization uses the \texttt{pathos} multiprocessing module to allow more robust serialization, and not the standard \texttt{multiprocessing} module. Assuming that the PRC, IRPC, and Floquet eigenfunctions have been computed, solving for $p_i^{(k)}$ up to order 4 with $M=4000$ takes approximately 45 minutes for the thalamic model on 8 cores. Solving $p_i^{(k)}$ for the CGL model only requires $M=200$ and takes roughly 5-10 minutes on two cores.
	
	\section{Thalamic Model}\label{a:thal}
	
	The remaining equations for the Thalamic model are
	\begin{align*}
		I_\text{L}(V) = g_\text{L} (V-E_L), &\quad I_{\text{Na}} = g_\text{Na} h m_\infty^3(V)(V-E_\text{Na}),\\
		I_\text{K} = 0.75 g_\text{K}(1-h)^4(V-E_\text{K}), &\quad I_\text{T} = g_\text{T} r p_\infty^2(V)  (V-E_\text{T}),
	\end{align*}
	and
	\begin{align*}
		a_h(V) = 0.128 \exp(-(V+46)/18),&\quad b_h(V) = 4/(1+\exp(-(V+23)/5)),\\
		m_\infty(V) = 1/(1+\exp(-(V+37)/7)),&\quad h_\infty(V) = 1/(1+\exp((V+41)/4)),\\
		r_\infty(V) = 1/(1+\exp((V+84)/4)),&\quad p_\infty(V) = 1/(1+\exp(-(V+60)/6.2)),\\
		\tau_h(V) = 1/(a_h(V)+b_h(V)),&\quad \tau_r(V) = 28 +\exp(-(V+25)/10.5).
	\end{align*}
	Please see Table \ref{tab:thal} for the parameters.
	
	\begin{table}
		\caption {Thalamic model parameter values} \label{tab:thal}
		\begin{center}
			\begin{tabular}{l|l}
				Parameter & Value\\
				\hline
				$C$& $1 \mu \text{F}/\text{cm}^2$\\
				$E_k$&$-90 \text{mV}$\\
				$E_{Na}$& $50 \text{mV}$\\
				$E_{t}$& $0 \text{mV}$\\
				$E_l$& $-70 \text{mV}$\\
				$E_{syn}$& $0 \text{mV}$\\
				$g_l$& $0.05 \text{mS}/\text{cm}^2$\\			
				$g_k$& $5 \text{mS}/\text{cm}^2$\\
				$g_{Na}$& $3 \text{mS}/\text{cm}^2$\\
				$g_\text{syn}$& $0-0.25\text{mS}/\text{cm}^2$\\
				$\alpha$ & 3\\
				$\beta$ & 2\\
				$\sigma_T$ & 0.8\\
				$V_{T}$&$-20 \text{mV}$\\
				$I_\text{app}$& $3.5 \mu \text{A}/\text{cm}^2$
			\end{tabular}
		\end{center}
	\end{table}
	
	\section*{Acknowledgments}
	The authors acknowledge support under National Institute of Health grant T32 NS007292 (YP) and National Science Foundation Grant No. CMMI-1933583 (DW).
	
	\bibliographystyle{plain}

\end{document}